\documentclass[sigconf]{acmart}
\providecommand\name{\textit{PigV$^2$}}
\AtBeginDocument{%
  \providecommand\BibTeX{{%
    \normalfont B\kern-0.5em{\scshape i\kern-0.25em b}\kern-0.8em\TeX}}}

\copyrightyear{2022} 
\acmYear{2022} 
\setcopyright{usgovmixed}\acmConference[SenSys '22]{The 20th ACM Conference on Embedded Networked Sensor Systems}{November 6--9, 2022}{Boston, MA, USA}
\acmBooktitle{The 20th ACM Conference on Embedded Networked Sensor Systems (SenSys '22), November 6--9, 2022, Boston, MA, USA}
\acmPrice{15.00}
\acmDOI{10.1145/3560905.3568416}
\acmISBN{978-1-4503-9886-2/22/11}


\begin{document}

\title{PigV$^2$: Monitoring Pig Vital Signs through Ground Vibrations Induced by Heartbeat and Respiration}


\author{Yiwen Dong}
\email{ywdong@stanford.edu}
\orcid{0000-0002-7877-1783}
\authornotemark[1]
\affiliation{%
  \institution{Stanford University}
  \streetaddress{450 Serra Mall}
  \city{Stanford}
  \state{California}
  \country{USA}
  \postcode{94305}
}
\author{Jesse R Codling}
\orcid{0000-0001-8355-7186}
\affiliation{%
  \institution{University of Michigan}
\city{Ann Arbor}
  \state{Michigan}
  \country{USA}}

\author{Gary Rohrer}
\orcid{0000-0002-8252-9308}
\affiliation{%
  \institution{USDA, ARS, U.S. Meat Animal Research Center, Clay Center}
  \country{USA}}
  
  \author{Jeremy Miles}
  \orcid{0000-0003-4765-8400}
\affiliation{%
  \institution{USDA, ARS, U.S. Meat Animal Research Center, Clay Center}
  \country{USA}}
  
\author{Sudhendu Sharma}
\orcid{0000-0003-1996-673X}
\affiliation{%
  \institution{University of Nebraska-Lincoln}
      \city{Lincoln}
  \state{Nebraska}
  \country{USA}} 

   \author{Tami Brown-Brandl}
 \orcid{0000-0002-0874-8035}
\affiliation{%
  \institution{University of Nebraska-Lincoln}
    \city{Lincoln}
  \state{Nebraska}
  \country{USA}}
  
\author{Pei Zhang}
\orcid{0000-0002-8512-1615}
\affiliation{%
  \institution{University of Michigan}
  \city{Ann Arbor}
  \state{Michigan}
  \country{USA}} 
  
 \author{Hae Young Noh}
 \orcid{0000-0002-7998-3657}
\affiliation{%
  \institution{Stanford University}
    \city{Stanford}
  \state{California}
  \country{USA}}

\renewcommand{\shortauthors}{Dong, et al.}

\begin{abstract}
Pig vital sign monitoring (e.g., estimating the heart rate (HR) and respiratory rate (RR)) is essential to understand the stress level of the sow and detect the onset of parturition. It helps to maximize peri-natal survival and improve animal well-being in swine production. The existing approach mainly relies on manual measurement, which is labor-intensive and only provides a few points of information. Other sensing modalities such as wearables and cameras are developed to enable more continuous measurement, but are still limited due to animal discomfort, data transfer, and storage challenges. In this paper, we introduce \name{}, the first system to monitor pig heart rate and respiratory rate through ground vibrations. Our approach leverages the insight that both heartbeat and respiration generate ground vibrations when the sow is lying on the floor. We infer vital information by sensing and analyzing these vibrations. The main challenge in developing \name{} is the overlap of vital- and non-vital-related information in the vibration signals, including pig movements, pig postures, pig-to-sensor distances, and so on. To address this issue, we first characterize their effects, extract their current status, and then reduce their impact by adaptively interpolating vital rates over multiple sensors. \name{} is evaluated through a real-world deployment with 30 pigs. It has 3.4\% and 8.3\% average errors in monitoring the HR and RR of the sows, respectively.
\end{abstract}



\maketitle

\section{Introduction}\label{sec:intro}

Monitoring pig vital signs (e.g., heart rate (HR) and respiratory rate (RR)) is important in providing pig health information for smart swine production~\cite{sipos2013physiological,carr2018pig}. Specifically, monitoring the sow's heart rate and respiratory rate in a farrowing pen enables predictions of stress levels and the onset of parturition~\cite{de1996rearing,jong2000effects,VONBORELL2007293,zaleski1993variables,randall1990induction}. This information helps the caretaker to determine which sow and when to assist in reducing neonatal mortality in swine production, improving animal well-being and production efficiency.

Traditional monitoring relies on farmers observing sow behavior, which is time-consuming, require professional training, and only produces sporadic results. In addition, manually measuring the pig's HR and RR causes stress to the animal and may expose them to health risks. Prior work adopted wearable sensors and contact-less cameras to collect such information~\cite{Macon2021,MARCHANTFORDE2004449,barbosa2019contactless,jorquera2020remotely,wang2021contactless}. Still, they face challenges with animal discomfort, battery life, data transmission, and insufficient bandwidth issues, which hampers the scalability. Prior work explored vibration-based monitoring for humans and animals, which utilizes vibration sensors attached to the floor to capture vibrations induced by the subjects~\cite{dong2020md,dong2023stranger}. Ground vibration sensing is contact-less to animals, wide-ranged, and has a $\sim$200$\times$ reduction of data transmission and processing requirement compared to the cameras~\cite{codlingSowPostureFeeding2022}. It has succeeded in pig weight monitoring and activity recognition, including nursing, posture changes, and growth patterns~\cite{bonde2021pignet,codling2021masshog,codlingSowPostureFeeding2022}.

In this paper, we present \name{} (\textbf{P}ig-\textbf{i}nduced \textbf{g}round \textbf{V}ibrations for \textbf{V}ital monitoring), which is the first system to achieve animal vital sign monitoring through ground vibration sensing. Our system leverages the knowledge that the sow's heartbeat and respiration exert forces onto the ground that generate vibrations. This insight allows us to monitor their rates by sensing and analyzing these vibration waves collected by low-cost, contact-less vibration sensors.

It is challenging to monitor pig vital signs through ground vibrations due to the mixture of vital- and non-vital-related information in the vibration signals. Specifically, vibration signals are influenced significantly by pig behaviors, including movements, different postures, locations, etc. In addition, these behaviors change over time and are hard to control because of the choices made by the individual pigs. Yet, they overshadow the vital-related information, leading to low signal-to-noise ratios (SNR) and bias in HR and RR estimations.  

To overcome this challenge, we characterize the vibration signal under the influence of various pig behaviors and develop a pig vital monitoring method with stable performances despite such changes. To achieve this, we first extract pig behaviors based on the characterization. Then, we reduce the impact of such behavior on vital monitoring by detecting and classifying them. Finally, we interpolate HR and RR when movements happen or the posture changes, and combine the results among multiple sensors. 

To evaluate our method, we conducted a real-world evaluation at a research farm in the U.S. with 30 pigs over multiple deployments. Our method has achieved an average of 3.4\% ($\pm$3.7 heartbeats per minute) and 8.3\% ($\pm$3.5 respirations per minute) error in HR and RR monitoring, respectively.

The contributions of this paper are:
\begin{itemize}
\item We develop the first animal vital sign monitoring system through ground vibration sensing, which enables continuous, contact-less, cost-efficient heart rate and respiratory rate monitoring.
\item We characterize how pig movements, postures, and pig-to-sensor distances affect the vibration signal for vital monitoring using physical insights of ground vibrations, and develop an algorithm to reduce such influence under changing pig behaviors.
\item We conduct a real-world evaluation with 30 pigs achieving an average of 3.4\% and 8.3\% errors in HR and RR monitoring, respectively.  
\end{itemize}

The remainder of the paper first presents the physics-informed characterization of how the pig behaviors affect the HR and RR monitoring (Section~\ref{sec:physical}), then introduces our method that monitors pig vital signs under various pig behavior influences (Section~\ref{sec:method}), and finally discuss the real-world experiment and evaluation results (Section~\ref{sec:eval}), followed by future work(Section~\ref{sec:futurework}) and conclusions (Section~\ref{sec:conclusion}).

\section{Physics-informed Characterization for Vibration-based Pig Vital Monitoring} \label{sec:physical}

\begin{figure}[t!]
\begin{center}
    \includegraphics[width=\linewidth]{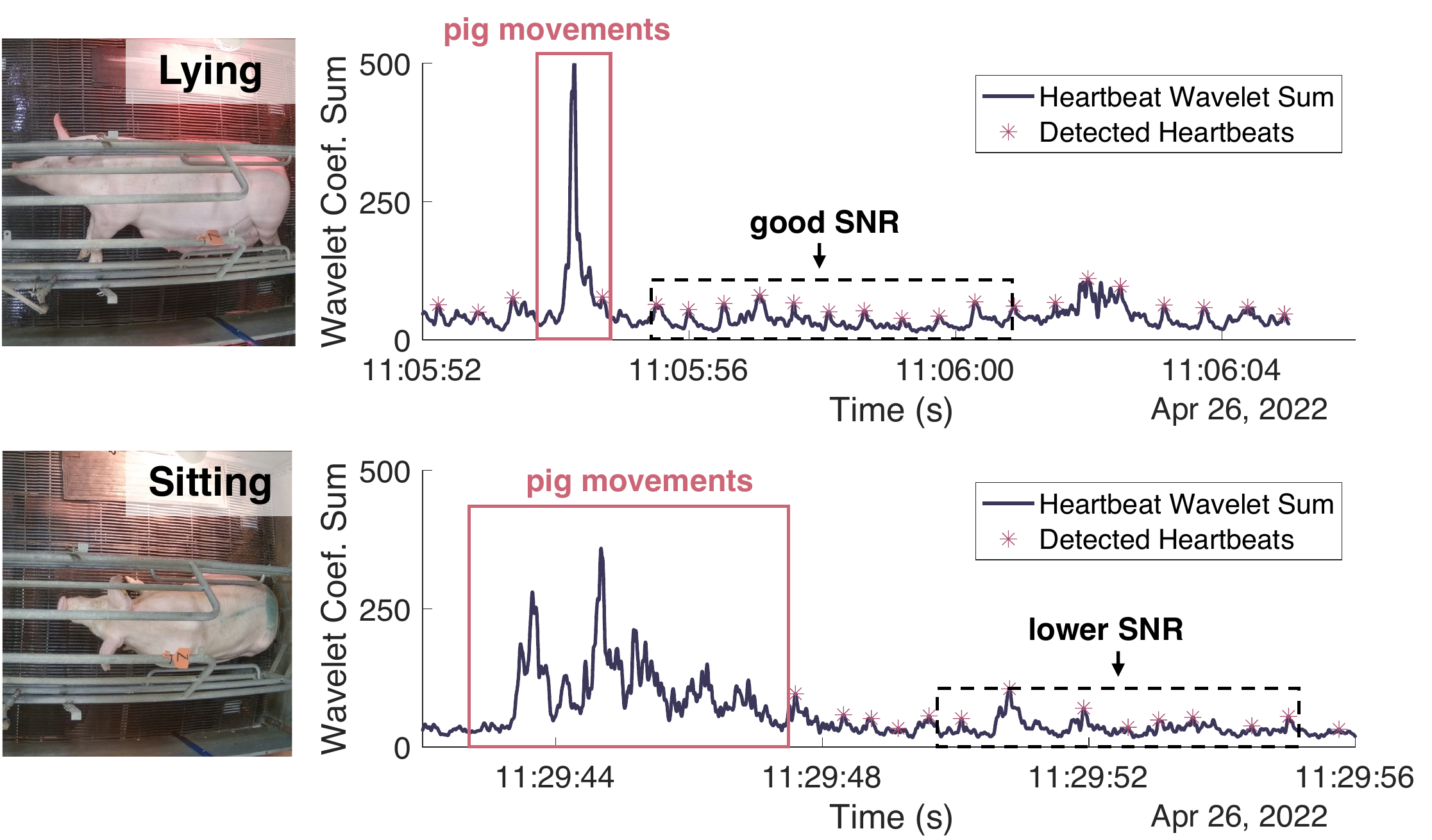}
  \caption{Samples of wavelet-transformed vibration signal during sitting and lying postures. The lying posture produces more significant impulses and better SNR than sitting. The high amplitude spikes (in pink boxes) correspond to the pig movements, which overshadow the heartbeats.}%
\label{fig:effects}
\vspace{-15pt}
\end{center}
\end{figure}

 \begin{figure*}[ht!]
\begin{center}
    \includegraphics[width=0.99\textwidth]{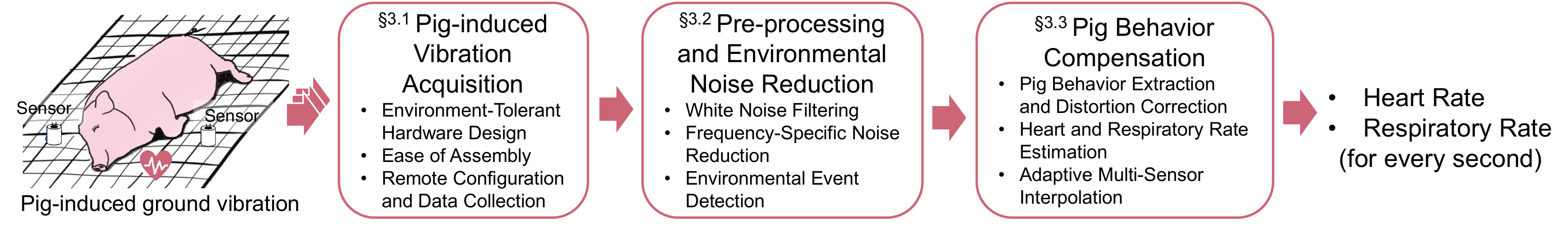}
  \caption{\name{} System Overview}%
\label{fig:system}
\vspace{-15pt}
\end{center}
\end{figure*}

The primary physical insight of our vibration-based pig vital monitoring is that ground vibrations are generated by the change of forces from the sow due to her heartbeat and respiration. When the sow is lying still, her body weight is assumed to be a constant force exerted on the floor. On top of that, her heartbeat and respiration induce slight body motion that exerts additional varying forces on the ground, which breaks the force equilibrium of the ground and results in vibrations to retain its equilibrium. These vibrations then propagate through the floor and are recorded by the vibration sensors attached to the bottom of the pig pen.

During the process mentioned above, the vibration signals are influenced by various pig behaviors, including 1) postures, 2) movements, and 3) pig-to-sensor distances. Moreover, these behaviors typically vary over time and are hard to control in real-world scenarios, leading to unstable performance and unreliable HR and RR estimations from the vibration signals. To overcome this issue, we first characterize the effect of each behavior to develop a system insensitive to such behavior changes.

\paragraph{\bf Effect of Pig Postures}
Different sow postures result in various signal patterns due to the difference in body motions and contact surfaces. First, the heartbeat and respiration cause unique motion in each part of the pig body, which pushes the floor in different ways. In addition, the weight transfer and the contact surface between the pig body and the floor may also vary when the posture changes. For example, heartbeat and respiration induce apparent motions on the sow's abdomen but much less on her hind-quarter and feet, so the lying posture produces more significant impulses in the vibration signals than standing and sitting. In addition, the lateral lying posture induces better SNR than sternal lying (see Figure~\ref{fig:effects}). This is because the weight transfer during lateral lying leads to a more stable gravity force and a larger contact surface.

\paragraph{\bf Effect of Pig Movements}
Movements from the sow induce significant changes in the force magnitudes exerted onto the ground, which typically overshadows the force changes due to heartbeat and respiration (see Figure~\ref{fig:effects}). Such movements happen for various reasons, including within-posture adjustments, posture changes, daily activities (e.g., ingestion and excretion), and so on. Based on our observations, posture adjustment induces minor signal changes among all, yet it still has a 5-10$\times$ increase in magnitude than the vibration induced by heartbeat.

\paragraph{\bf Effect of Pig-to-Sensor Distances}
The changing distance between the pig and the sensor affects the recorded vibration signals because the signals are distorted during the vibration wave propagation. Such distortion is mainly due to the effect of 1) wave attenuation and 2) wave dispersion. As the wave attenuates over longer distances, the magnitude of vital sign-induced vibrations decreases exponentially, so sensors outside the sensing range do not capture the vital signs (the sensing range is around 2 meters based on preliminary testing). In addition, since waves of different frequencies travel at various velocities, the arrival time of their peaks tends to misalign as the sensing distance increases, leading to multiple smaller peaks per heartbeat.

\section{Pig Vital Sign Monitoring through Ground Vibrations}\label{sec:method}

\name{} consists of three modules: 1) pig-induced vibration acquisition, 2) pre-processing and environmental noise reduction, and 3) pig behavior compensation (as described in Figure~\ref{fig:system}).

\subsection{Pig-Induced Vibration Acquisition}

\name{} acquires vibration data from a robust network of geophone sensors mounted below the pigs' pens similar to past works~\cite{codling2021masshog,codlingSowPostureFeeding2022,bonde2021pignet}.
The network is arranged into tiers to overcome the tradeoffs of reliability and sensing performance~\cite{codling2021masshog}.
At the lowest tier are low-power geophone nodes which convert the vertical movement velocity of the pigpen floor to a stream of data.
The second tier aggregates these streams and forwards them to the final tier.
This tier also manages the nodes based on remote configuration.
Off-\-site resources comprise the final tier where the data processing occurs.

In its current iteration, the sensor nodes are low-power in terms of processing capability, but not in terms of electricity consumption.
Our sensors are optimized to survive the harsh farm environment (e.g., water flushing, animal activities) and ease of assembly first.
While simpler, low-capability devices improve robustness through redundancy, their power consumption has thus far been less critical.
Given the nature of their sensing task, however, these sensors could be made to function without external power.
Minimal hardware adjustment can readily trade some configuration responsiveness for significant gains in power consumption for off-grid use.


Remote management is a key facilitator enabling deployment by non-engineers.
In \name{}, the aggregator tier connects to the sensor nodes via a wireless link, but can also be accessed remotely through the same way it sends data off to the processing tier.
Once supplied with an internet connection, a remote specialist can perform software-level setup as local staff mounts the hardware during deployment.
As data are collected, the sensors' gain, sampling, and metadata can be manipulated as needed by sending remote configuration changes to the aggregator, even though the physical devices are inaccessible when pigs are present.
This configurable network design improves reliability in the farm environment through ease of deployment by non-engineers and its ability to adapt to unknown and changing conditions.

\subsection{Data Pre-processing and Environmental Noise Reduction}

The vibration signals are pre-processed through filters, an environmental event detection algorithm, and wavelet decomposition to reduce various environmental noises. 

The signals are first applied with a Wiener filter to reduce the white noise. Then, we process the signal through a band-pass filter and select the 0 to 200 Hz band for further analysis. This step allows us to reduce the high-frequency noises from the electrical and mechanical components in the farm environment (i.e., fans, air conditioners) while preserving the pig's vital-related information of lower frequencies. 

Noises induced by environmental events (e.g., water flushing, staff walking, tractors passing by) are detected and removed by matching the signal with pre-existing templates of those events. To achieve this, we first crop the signals based on the scheduled start and end times of regular staff check-in and cleaning. Then, within these cropped signals, we apply a sliding window of 10 seconds with 50\% of overlap to divide the continuous signal into segments. The 10-second window was chosen based on the maximum duration of those events. Each segment is compared with the existing templates by taking the cross-correlation for alignment and cosine distance for similarity comparison. Windows with similarity scores higher than the threshold are marked as environmental noises.

Finally, we apply wave transform to the processed signal. During the wavelet transform, we choose the generalized Morse wavelet as the basis function because it has a single peak for each heartbeat-/respiration-induced impulse, allowing HR and RR estimation by counting the number of peaks per minute in the transformed signal.

\subsection{Pig Behavior Compensation}

In this section, we monitor pig HR and RR according to the changing pig behaviors (including pig postures, movements, and pig-to-sensor distances). First, we extract the pig behaviors from the vibration signals over time and remove the non-vital-related signal segments. Then, we develop a method based on wavelet decomposition and peak detection to estimate HR and RR. Finally, we combine the results from multiple sensors in each pen for adaptive interpolation when non-vital-related pig behavior happens. 

\subsubsection{Pig Behavior Extraction and Distortion Correction}
We extract the pig behaviors from the signals in order to reduce the bias in estimation and improve the consistency in performance over time. As discussed in Section~\ref{sec:physical}, these behaviors include 1) pig postures, 2) pig movements, and 3) pig-to-sensor distances. 

\begin{figure}[t!]
\begin{center}
    \includegraphics[width=\linewidth]{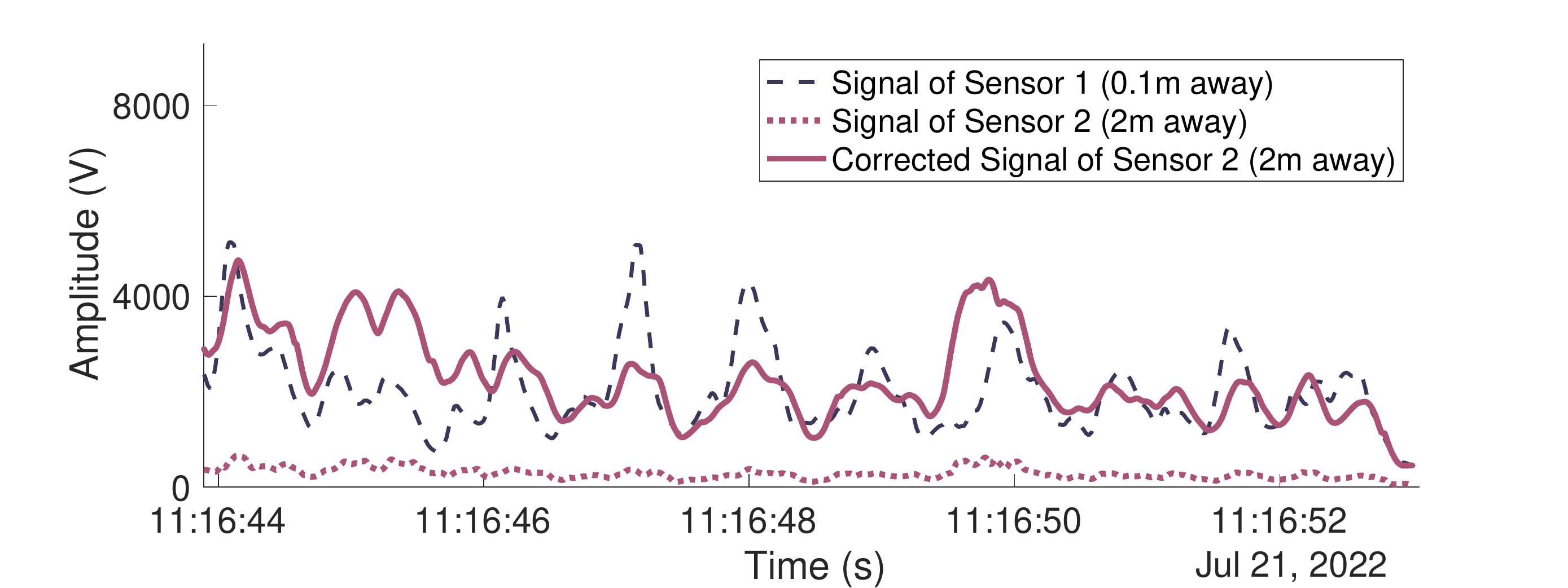}
  \caption{Signals from a sensor away from the pig (pink dashed line) are corrected based on the dispersion and attenuation effect during wave propagation from the pig to the sensor location. The peaks from the corrected signal (pink solid line) align well with the signals from the sensor right underneath the pig (black dashed line).}%
\label{fig:reverse}
\vspace{-10pt}
\end{center}
\end{figure}

 \paragraph{\bf Pig Posture Classification}
The pig's posture significantly affects the vital-related signal-to-noise ratio (SNR). Therefore, classifying the pig posture helps improve the estimation accuracy by removing the segments below our SNR requirements. In \name{}, pig posture is predicted as either lying, sitting/kneeling, and standing, using the random forest classifier trained in our prior study~\cite{codlingSowPostureFeeding2022,bonde2021pignet}. A preliminary study found that pigs lie more than 80\% of the time. Meanwhile, sitting and standing do not produce sufficient SNR for vital sign monitoring, as discussed in Section~\ref{sec:physical}. Based on the above evidence, we prioritize the signals from the lying posture (including both lateral and sternal lying) and cut out the signals from the other postures. The cut out signal duration will be compensated by interpolation in Section~\ref{subsec:fusion}.

\paragraph{\bf Pig Movement Detection}
 Pig movements are detected based on an anomaly detection algorithm in order to remove the non-vital-related signal segments that bias the vital rate estimation. First, the algorithm takes a 10-second vibration signal when the sow is lying still with a regular heartbeat and respiration as a reference. Then, a 1-second sliding window is applied to the signal to detect the pig movements: any window with a mean larger than three standard deviations from the reference signal (i.e., out of 99.7\% confidence interval) is detected as a window containing pig movements. Finally, signals within these windows are removed and will be interpolated using adjacent vital rate estimations and/or readings from the nearby sensors, which will be discussed in Section~\ref{subsec:fusion}.

\paragraph{\bf Pig-to-Sensor Signal Distortion Correction}

The vibration signals are distorted during wave propagation. As discussed in Section~\ref{sec:physical}, such distortion is reduced in two aspects: wave attenuation and wave dispersion. 

We first compute the attenuation coefficient $\alpha$ during preliminary testing with at least two sensors to correct the wave attenuation effect. Assuming a fixed attenuation equation within the entire pen, we solve the inverse problem and revert the signal energy to match the amplitudes measured by the sensor right below the pig. 
\begin{equation}
    S_{loc2} = S_{loc1} e^{-\alpha fd} 
\end{equation}
where $S$ is the vibration signal amplitude, $f$ is the frequency component, and $d$ is the pig-to-sensor distance.

The dispersion effect is challenging to correct due to the unknown wave propagation velocities when it travels through the floor. Therefore, instead of reverting the signal to the pig location, we choose to mitigate the dispersion effect to improve the vital monitoring accuracy. Since wave dispersion results in misaligned signal peaks (as mentioned in Section~\ref{sec:physical}), we re-align those peaks by taking the moving average along the pre-processed signal based on an approximated time lag for each sensing location, as shown in Figure~\ref{fig:reverse}. For example, the vibration wave typically travels around 100-200 m/s according to previous studies~\cite{mirshekari2018occupant}, a 0.05 second (i.e., 25 samples) moving average is chosen when the pig-to-sensor distance is 5 meters.

\subsubsection{Heart Rate and Respiratory Rate Estimation}
After reducing the effect of pig behaviors, we estimate the HR and RR for each sensor by counting the number of detected heartbeats and respirations within a minute. 

\begin{figure}[t!]
\begin{center}
    \includegraphics[width=\linewidth]{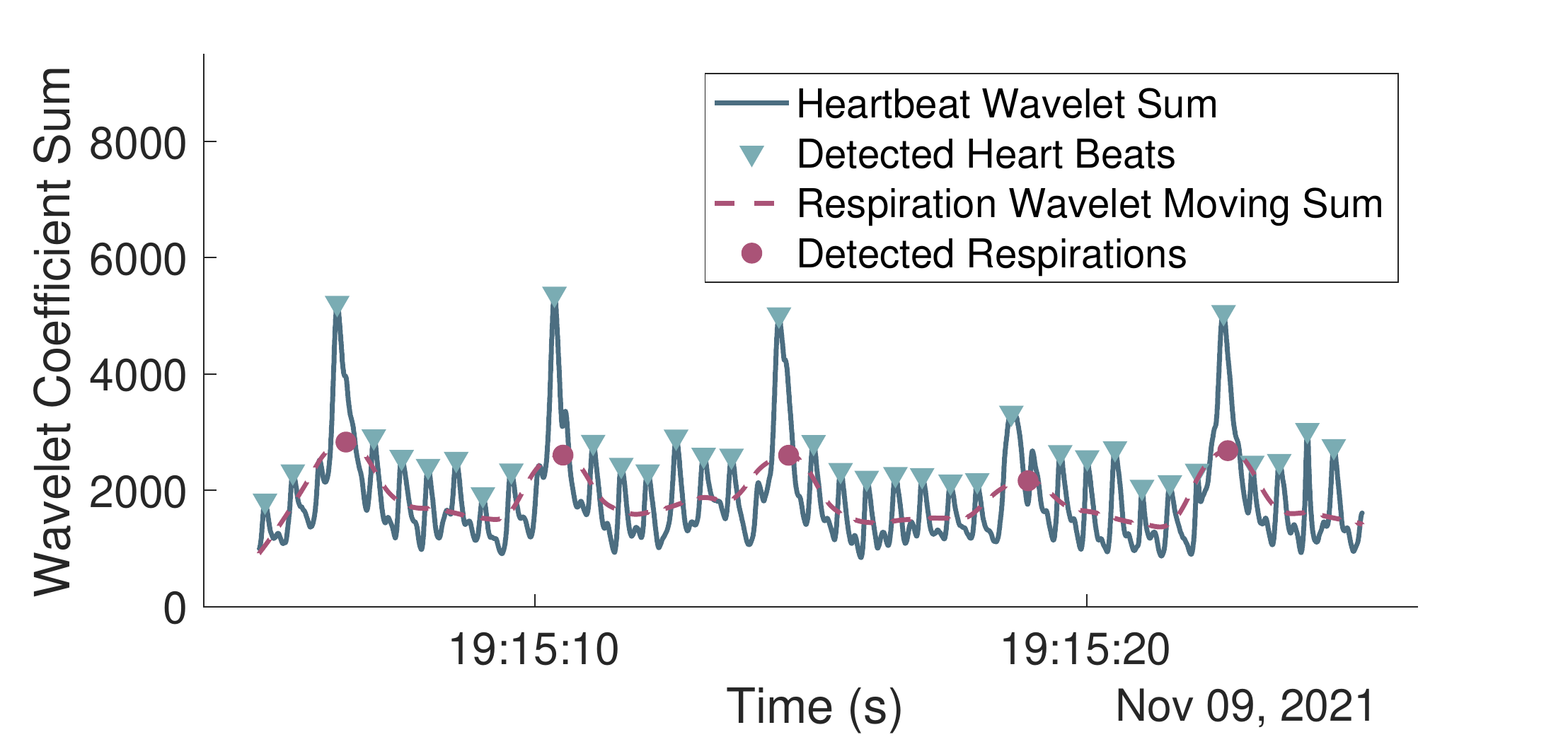}
  \caption{Heartbeats (green triangles) and respirations (pink circles) are detected in the vibration signals when the sow is lying still.}%
\label{fig:signal}
\vspace{-10pt}
\end{center}
\end{figure}

The heartbeats are detected through peak picking over the sum of wavelet coefficients from 10 to 100 Hz (see the solid green line in Figure~\ref{fig:signal}). The range is chosen based on the typical heartbeat-induced vibration frequency range (0-100 Hz) and the sensitivity range of the sensors ($\ge$10 Hz). Since the HR of the sow typically ranges from 60 to 120 beats per minute, we set the minimum peak prominence as 0.5 seconds to avoid the false detection of the adjacent lower peak caused by diastole.

Unlike the impulse generated by the heartbeat, respiration induces a slower change in forces when the sow's body presses onto the ground, indirectly affecting the energy trend of the heartbeat impulses. Therefore, respiration is detected by pick peaking over the moving sum of the wavelet coefficients to capture the energy trend of the impulses (see the dashed pink line in Figure~\ref{fig:signal}). Since the RR of the sow typically ranges from 10 to 60 breaths per minute, we set the minimum peak prominence as 1 second to reduce the false positive rate caused by strong heartbeats.

\subsubsection{Adaptive Multi-Sensor Interpolation}\label{subsec:fusion}
To monitor pig vital signs continuously with reliable HR and RR estimations, we interpolate the results over time by adaptively combining the results from multiple sensors underneath the same pen. 

We first conduct interpolation on each sensor during the removed signals when the sow is in sitting/standing postures or in motion. An existing study suggests that the change of postures often leads to an immediate and significant increase in HR and then decreases gradually after the sow returns to the resting posture~\cite{borst1982mechanisms}. Therefore, we choose spline interpolation to model such a mechanism, which captures the sudden increase and gradual decrease more effectively than a linear function.

After the interpolation, we combine the results from multiple sensors to ensure continuous and accurate monitoring of vital signs. To avoid system interruptions due to seldom sensor connection issues, we continuously monitor the functional status of each sensor and conduct analysis only on the active sensors. There is a tradeoff when combining various sensors. While the sensors closer to the sow typically have a better SNR on vital signs, they are more sensitive to the pig's movements. Therefore, the sensors closer to the pig typically require a longer duration of interpolation, which hampers vital monitoring accuracy. Therefore, we compute the weighted average of multiple sensors to estimate the vital rates. The weight $w_s$ based on the interpolation duration $t_{s}^{int}$ and the distance from the pig $d_s$, computed as follows:
\begin{equation}
    w_{s} = \frac{T-t_{s}^{int}}{T} e^{-d_s}
\end{equation}
where $T$ is the length of time for each round of peak detection (i.e., 60 seconds in this case). This formula is developed based on the exponential decrease of SNR w.r.t. the sensing distance and the linear decrease of vital-related information as the interpolation duration increases. 

\section{Real-world Evaluation} \label{sec:eval}
We evaluate our system through multiple field deployments on a research farm in the USA. In this section, we first discuss our experiment setup, then show the system performance in terms of prediction accuracy, and finally discuss and demonstrate its robustness to the changing pig behaviors.

\subsection{Field Experiment Implementation}
We conducted multiple deployments for \name{} on the U.S. Meat Animal Research Center in Nebraska, USA,
with 30 pigs. Data collection was performed in accordance with federal and institutional regulations regarding proper animal care practices and was approved by the U.S. Meat Animal Research Center's Institutional Animal Care and Use Committee as $EO\#143.0.$ and $EO\#171.0$.

\begin{figure}[t!]
\begin{center}
    \includegraphics[width=\linewidth]{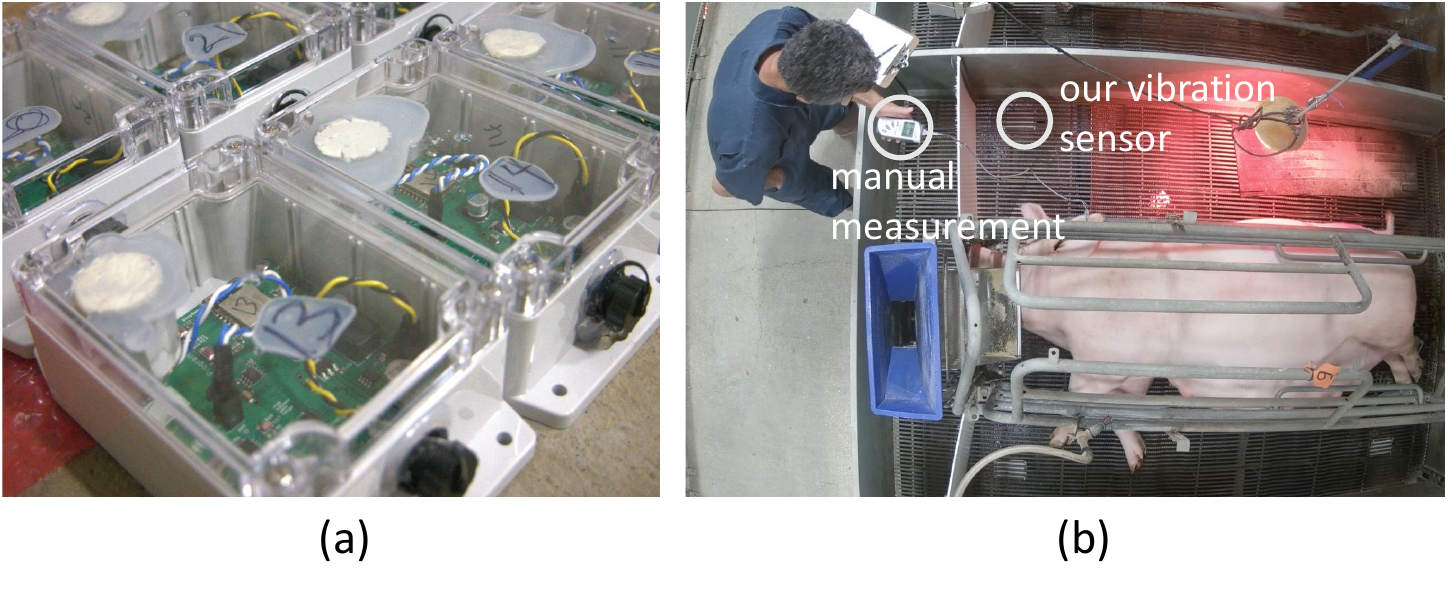}
  \caption{Experiment setup: a) our vibration sensor in waterproof boxes, b) manual measurement for ground truths and our contact-less vibration sensor attached to the bottom of the floor.}%
\label{fig:deployment}
\vspace{-10pt}
\end{center}
\end{figure}

We installed 30 geophone sensors over eight pens, in which three pens are deployed with 5 sensors, and the other five are deployed with 3 sensors. The sampling frequency is 500 Hz.
All sensors are concealed in plastic boxes with a watertight enclosure to prevent damage from high-pressure water during cleaning and excretion from the sow during operation (see Figure~\ref{fig:deployment}a). The sensors were all installed underneath the pen, fixed with multiple zip-ties to ensure firm coupling with the floor structure.  

The ground truth was collected by our animal scientist on the farm (see Figure~\ref{fig:deployment}b). Video cameras are also installed above the pen to provide information on pig postures and movements. The heart rate was collected using Edan VE-H100B, a veterinary pulse oximeter (Pulse Ox) that measures the pulse rate through an ear clip. 
The device has an error of $\pm$2 beats per minute ($\sim$2\%). The RR was counted manually by watching the pig's flank (i.e., abdomen) raise and lower, which is estimated to have an accuracy of $\pm$3 respirations per minute ($\sim$7\%) by comparing the counts from two individuals. To mitigate the risk during data collection, each sow is monitored for 1-2 minutes. The average heart and respiratory rates per minute are 98 and 35, respectively. The number tends to be higher than normal because HR and RR usually increase when a human approaches the sow and takes measurements~\cite{marchant2001vocalisations}.

\subsection{System Performance}
\begin{figure}[t!]
\begin{center}
    \includegraphics[width=\linewidth]{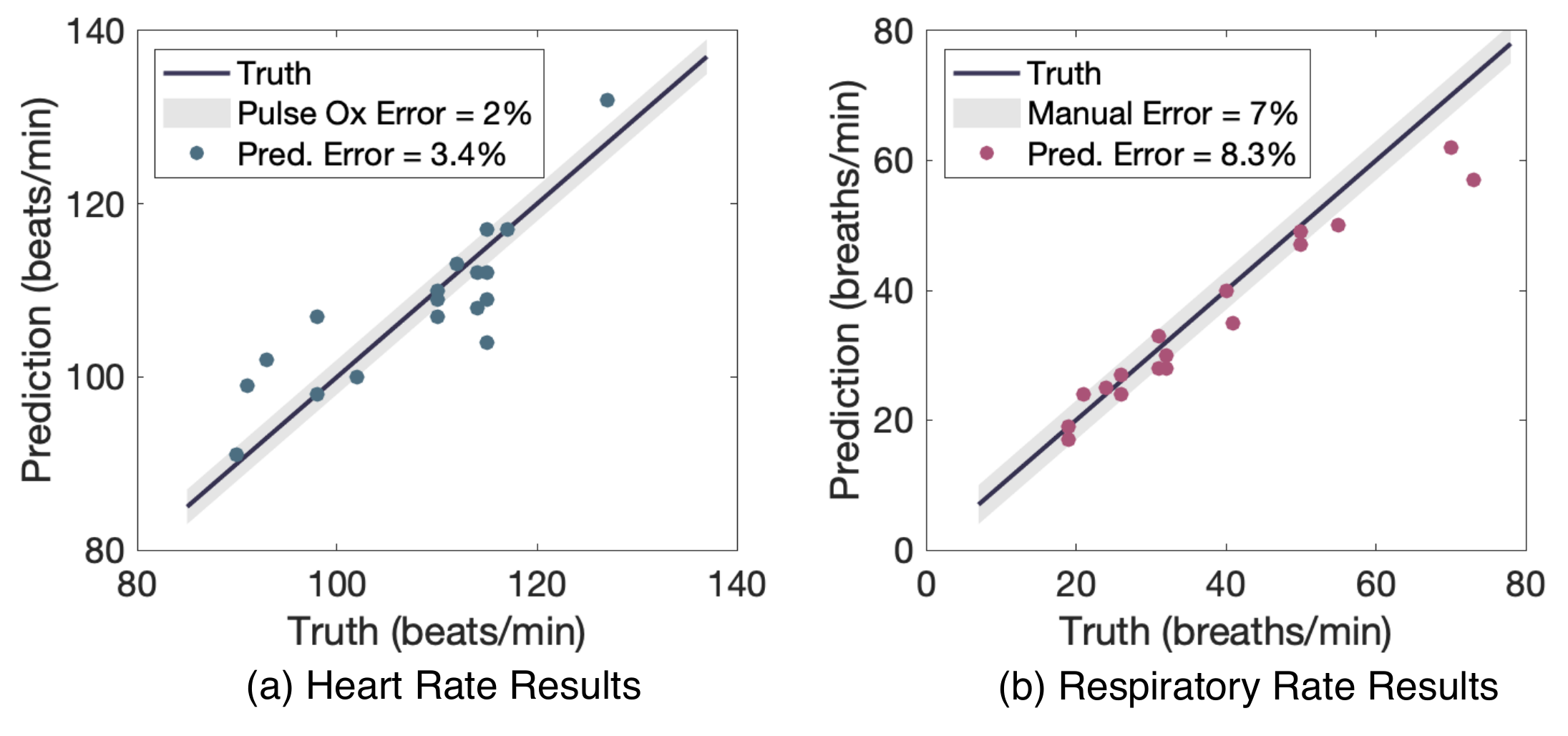}
  \caption{Overall performance of our \name{} system: a) the HR prediction error is 3.4\% ($\pm$3.7 beats per minute), b) the RR prediction error is 8.3\% ($\pm$3.5 breaths per minute). Both errors are comparable to the Pulse Ox and manual counting error, which is 2\% and 7\%, respectively.}%
\label{fig:overall}
\end{center}
\end{figure}

Overall, \name{} has an average error of 3.4\% and 8.3\% for HR and RR, respectively. As illustrated in Figure~\ref{fig:overall}, the error rate is comparable to the accuracy of Pulse Ox and manual RR observations, which has an error rate of 2\% and 7\% for HR and RR, respectively. The accuracy is sufficient for predicting the stress level and onset of parturition in pigs, which typically has a more than a 7-20\% increase in HR and a 20-50\% increase in RR based on previous studies~\cite{de1996rearing,randall1990induction,smith1956respiration}.

\begin{figure}[t!]
\begin{center}
    \includegraphics[width=\linewidth]{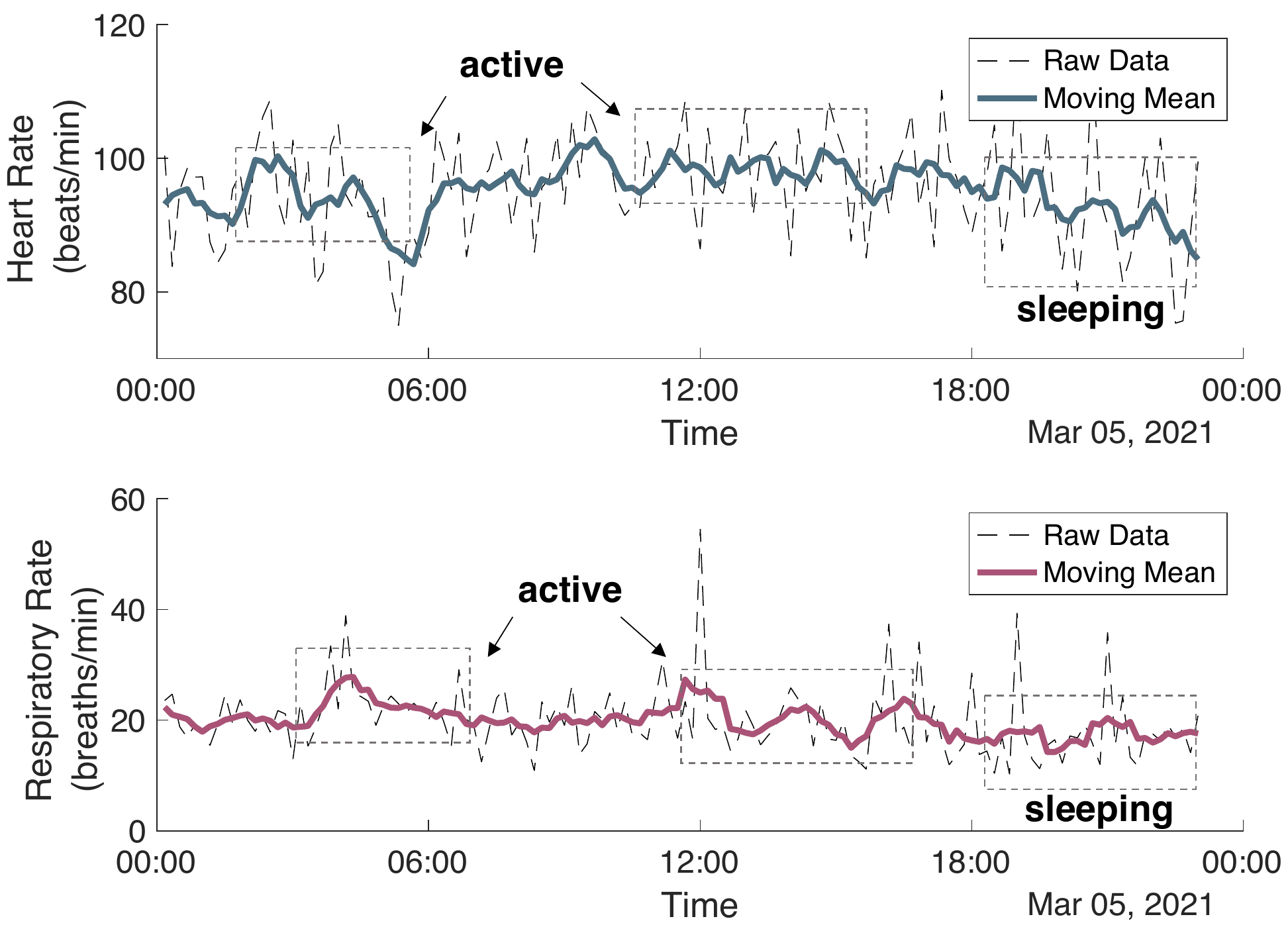}
  \caption{Daily trends of HR and RR share similar trends during active and sleeping times. The variation during sleeping may indicate the REM/non-REM sleep stages.
}%
\label{fig:dailyvital}
\vspace{-10pt}
\end{center}
\end{figure}

\subsubsection{Daily Patterns}

We also observe the daily patterns of pig vital signs to understand the sow's behavior changes over time. Figure~\ref{fig:dailyvital} shows the HR and RR daily trends after pig behavior compensation (with interpolation when the pig is moving). It appears that HR and RR have similar trends. They fluctuate more and have a higher rate in the morning (around 5 am) and the afternoon (4 pm) when they tend to be more active. In addition, a decreasing trend is observed when the lights are turned off (i.e., from 6 pm to 12 pm) because the sow primarily sleeps during that time. 

It is worth noting that there are regular fluctuations during the sow's sleeping, which may indicate the sleeping stages of the sow. Based on the existing study, the non-REM stage leads to slower breaths and heartbeats, while the REM stage (when dreams happen) leads to higher HR and RR~\cite{vzemaityte1984heart}. We will explore this trend in future work.

\subsubsection{Robustness Assessment on Pig Behavior Changes}
We evaluate the sensitivity of \name{} with various pig behavior changes to understand the system robustness and generalizability towards these changes. As discussed in Section~\ref{sec:physical} and~\ref{sec:method}, these context include pig postures, movements, and pig-to-sensor distances.

\paragraph{\bf Discussion on Pig Posture Robustness}

\begin{figure}[t!]
\begin{center}
\includegraphics[width=\linewidth]{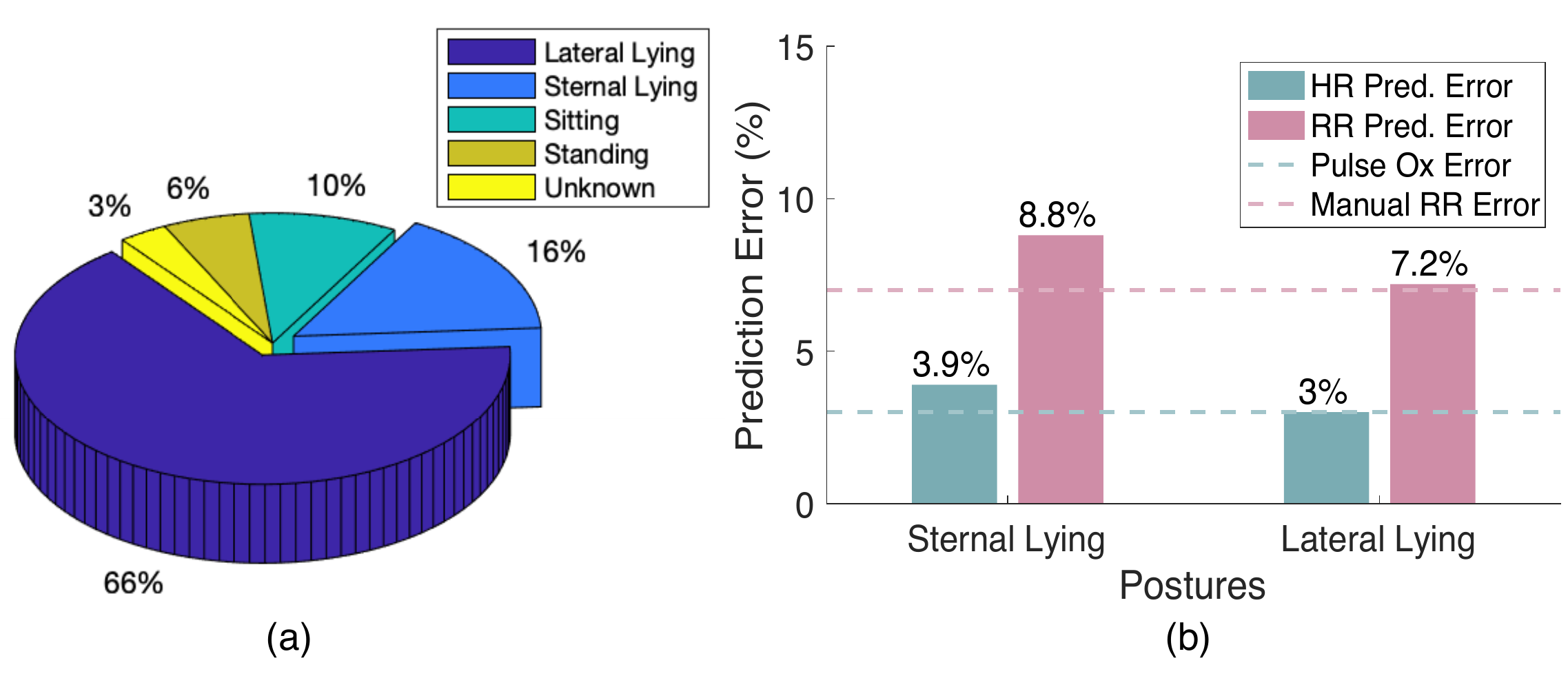}
  \caption{\name{} adapts to different postures. a) percentage distribution over different postures, where lying takes 82\% of time; b) the error of \name{} in lateral lying posture is slightly smaller than sternal lying, and both are comparable to the Pulse Ox and manual RR error.}%
\label{fig:result1}
\vspace{-10pt}
\end{center}
\end{figure}

We observe the posture distribution based on the video data to understand the percentage of lying, a posture that captures the most vital-related information. As shown in Figure~\ref{fig:result1}a, the lying posture covers 82\% of the time during our deployment, which means most signals contain vital information. Figure~\ref{fig:result1}b shows the system robustness over lateral and sternal postures. While sternal posture leads to slightly lower accuracy, it is still comparable to the ground truth accuracy.

\paragraph{\bf Effectiveness of Interpolation during Pig Movements}
We evaluate the effectiveness of interpolation by comparing it with the baseline method without compensating for the pig behaviors. Results show that the prediction error is reduced by 3.6$\times$ for HR and 2$\times$ for RR. Our approach effectively reduces the false positives caused by movement-induced impulses and missing heartbeats that are overshadowed by the movements.

\begin{figure}[t!]
\begin{center}
\includegraphics[width=0.9\linewidth]{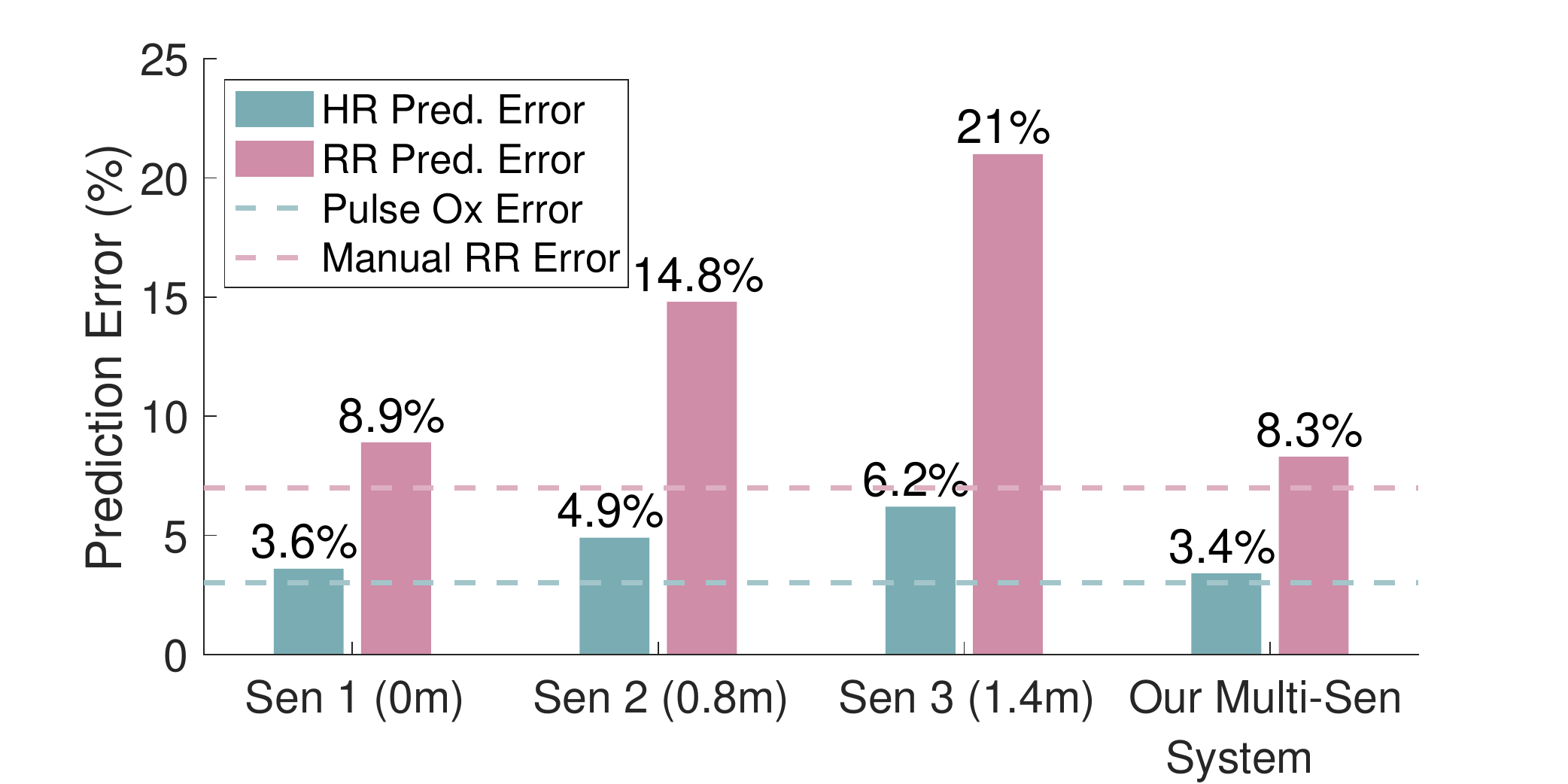}
  \caption{Effectiveness of pig-to-sensor distance correction and multi-sensor fusion. While the accuracy decreases as the distance increases, \name{} achieves a lower error than all of these sensors by correcting the distortion effect during pig-to-sensor wave propagation and combining results from multiple sensors.}%
\label{fig:result2}
\vspace{-15pt}
\end{center}
\end{figure}

\paragraph{\bf Effectiveness of Pig-to-Sensor Distortion Correction and Sensor Fusion}
We compared the prediction errors across various sensors to show the effectiveness of adaptive multi-sensor fusion. As described in Figure~\ref{fig:result2}, the sensor right underneath the sow produces the lowest error. As the pig-to-sensor distance increases, the prediction accuracy decreases because the heartbeat- and respiration-induced signals distort significantly during wave propagation, resulting in low SNR. Our system reduces the error by distance correction and sensor fusion, which leads to lower error than any of the sensors' predictions.
\section{Future Work}\label{sec:futurework}
There are many directions to explore in the future, given the rich information inherent in the pig-induced ground vibration data. Specifically, we identified three important aspects based on our observations and understandings developed through this study.

\textbf{Stress level and sleep stage monitoring:} HR and RR are good indicators of pigs' stress levels and sleeping stages~\cite{marchant2001vocalisations}. In this study, we observed changes in both rates when the farm staff approached the sow and when the sow was sleeping. Further study on the relationship between the vital rates and the sow's stress levels and sleeping stages can be helpful for caretakers to provide timely help to ensure their well-being.

\textbf{Onset of parturition prediction:} An increase in HR and RR is also an important indicator for parturition. Our prior study has observed increasing signal energy before the sow gives birth. We will further explore vibration-based methods to predict the parturition time. 

\textbf{Cardiac cycle disorder detection:} We observed that the heartbeat-induced vibrations typically have two peaks - a primary peak followed by a secondary peak, which may correspond to the systole and diastole phases of the cardiac cycle. By extracting the prominence and the relative amplitudes of these two peaks, we may detect disorders affecting the cardiac cycle.

In addition, we will continue exploring vital sign monitoring when there are multiple animals in the same pen and analyze the effect of different types of pens.

\section{Conclusions}\label{sec:conclusion}
In conclusion, we introduce \name{}, the first system to monitor pig HR and RR through ground vibrations. Our approach leverages the physical insight that heartbeat and respiration induce ground vibrations through the sow's body to monitor these vital signs. The main challenge in developing \name{} is the mixture of vital-related and non-vital-related information in the vibration signals. To overcome this challenge, we reduce the influence of non-vital-related information by first detecting and characterizing various pig behaviors and then interpolating the vital rates during these periods among multiple sensors. \name{} is evaluated through a real-world experiment with 30 pigs, with 3.4\% and 8.3\% average error in monitoring their HR and RR, respectively.


\begin{acks}
This research was supported in part by the National Science Foundation (NSF-CMMI-2026699) and Cisco, Inc. The USDA prohibits discrimination in all its programs and activities on the basis of race, color, national origin, age, disability, and where applicable, sex, marital status, familial status, parental status, religion, sexual orientation, genetic information, political beliefs, reprisal, or because all or part of an individual's income is derived from any public assistance program (Not all prohibited bases apply to all programs.). Persons with disabilities who require alternative means for communication of program information (Braille, large print, audiotape, etc.) should contact USDA's TARGET Center at (202) 720-2600 (voice and TDD). USDA is an equal opportunity employer.
\end{acks}

\bibliographystyle{ACM-Reference-Format}
\bibliography{reference}


\begin{thebibliography}{22}


\ifx \showCODEN    \undefined \def \showCODEN     #1{\unskip}     \fi
\ifx \showDOI      \undefined \def \showDOI       #1{#1}\fi
\ifx \showISBNx    \undefined \def \showISBNx     #1{\unskip}     \fi
\ifx \showISBNxiii \undefined \def \showISBNxiii  #1{\unskip}     \fi
\ifx \showISSN     \undefined \def \showISSN      #1{\unskip}     \fi
\ifx \showLCCN     \undefined \def \showLCCN      #1{\unskip}     \fi
\ifx \shownote     \undefined \def \shownote      #1{#1}          \fi
\ifx \showarticletitle \undefined \def \showarticletitle #1{#1}   \fi
\ifx \showURL      \undefined \def \showURL       {\relax}        \fi
\providecommand\bibfield[2]{#2}
\providecommand\bibinfo[2]{#2}
\providecommand\natexlab[1]{#1}
\providecommand\showeprint[2][]{arXiv:#2}

\bibitem[Barbosa~Pereira et~al\mbox{.}(2019)]%
        {barbosa2019contactless}
\bibfield{author}{\bibinfo{person}{Carina Barbosa~Pereira},
  \bibinfo{person}{Henriette Dohmeier}, \bibinfo{person}{Janosch Kunczik},
  \bibinfo{person}{Nadine Hochhausen}, \bibinfo{person}{Ren{\'e} Tolba}, {and}
  \bibinfo{person}{Michael Czaplik}.} \bibinfo{year}{2019}\natexlab{}.
\newblock \showarticletitle{Contactless monitoring of heart and respiratory
  rate in anesthetized pigs using infrared thermography}.
\newblock \bibinfo{journal}{\emph{Plos one}} \bibinfo{volume}{14},
  \bibinfo{number}{11} (\bibinfo{year}{2019}), \bibinfo{pages}{e0224747}.
\newblock


\bibitem[Bonde et~al\mbox{.}(2021)]%
        {bonde2021pignet}
\bibfield{author}{\bibinfo{person}{Amelie Bonde}, \bibinfo{person}{Jesse~R
  Codling}, \bibinfo{person}{Kanittha Naruethep}, \bibinfo{person}{Yiwen Dong},
  \bibinfo{person}{Wachirawich Siripaktanakon}, \bibinfo{person}{Sripong
  Ariyadech}, \bibinfo{person}{Akkarit Sangpetch}, \bibinfo{person}{Orathai
  Sangpetch}, \bibinfo{person}{Shijia Pan}, \bibinfo{person}{Hae~Young Noh},
  {et~al\mbox{.}}} \bibinfo{year}{2021}\natexlab{}.
\newblock \showarticletitle{PigNet: Failure-Tolerant Pig Activity Monitoring
  System Using Structural Vibration}. In \bibinfo{booktitle}{\emph{Proceedings
  of the 20th International Conference on Information Processing in Sensor
  Networks (co-located with CPS-IoT Week 2021)}}. \bibinfo{pages}{328--340}.
\newblock


\bibitem[Borst et~al\mbox{.}(1982)]%
        {borst1982mechanisms}
\bibfield{author}{\bibinfo{person}{C Borst}, \bibinfo{person}{W Wieling},
  \bibinfo{person}{JF Van~Brederode}, \bibinfo{person}{A Hond},
  \bibinfo{person}{LG De~Rijk}, {and} \bibinfo{person}{AJ Dunning}.}
  \bibinfo{year}{1982}\natexlab{}.
\newblock \showarticletitle{Mechanisms of initial heart rate response to
  postural change}.
\newblock \bibinfo{journal}{\emph{American Journal of Physiology-Heart and
  Circulatory Physiology}} \bibinfo{volume}{243}, \bibinfo{number}{5}
  (\bibinfo{year}{1982}), \bibinfo{pages}{H676--H681}.
\newblock


\bibitem[Carr et~al\mbox{.}(2018)]%
        {carr2018pig}
\bibfield{author}{\bibinfo{person}{John Carr}, \bibinfo{person}{Shih-Ping
  Chen}, \bibinfo{person}{Joseph~F Connor}, \bibinfo{person}{Roy Kirkwood},
  {and} \bibinfo{person}{Joaquim Segal{\'e}s}.}
  \bibinfo{year}{2018}\natexlab{}.
\newblock \bibinfo{booktitle}{\emph{Pig health}}.
\newblock \bibinfo{publisher}{CRC Press}.
\newblock


\bibitem[Codling et~al\mbox{.}(2021)]%
        {codling2021masshog}
\bibfield{author}{\bibinfo{person}{Jesse~R Codling}, \bibinfo{person}{Amelie
  Bonde}, \bibinfo{person}{Yiwen Dong}, \bibinfo{person}{Siyi Cao},
  \bibinfo{person}{Akkarit Sangpetch}, \bibinfo{person}{Orathai Sangpetch},
  \bibinfo{person}{Hae~Young Noh}, {and} \bibinfo{person}{Pei Zhang}.}
  \bibinfo{year}{2021}\natexlab{}.
\newblock \showarticletitle{MassHog: Weight-Sensitive Occupant Monitoring for
  Pig Pens using Actuated Structural Vibrations}. In
  \bibinfo{booktitle}{\emph{Adjunct Proceedings of the 2021 ACM International
  Joint Conference on Pervasive and Ubiquitous Computing and Proceedings of the
  2021 ACM International Symposium on Wearable Computers}}.
  \bibinfo{pages}{600--605}.
\newblock


\bibitem[Codling et~al\mbox{.}({[n.\,d.]})]%
        {codlingSowPostureFeeding2022}
\bibfield{author}{\bibinfo{person}{Jesse~R Codling}, \bibinfo{person}{Yiwen
  Dong}, \bibinfo{person}{Amelie Bonde}, \bibinfo{person}{Adeola Bannis},
  \bibinfo{person}{Asya Macon}, \bibinfo{person}{Gary Rohrer},
  \bibinfo{person}{Jeremy Miles}, \bibinfo{person}{Sudhendu Sharma},
  \bibinfo{person}{Tami Brown-Brandl}, \bibinfo{person}{Hae~Young Noh}, {and}
  \bibinfo{person}{Pei Zhang}.} \bibinfo{year}{[n.\,d.]}\natexlab{}.
\newblock \showarticletitle{Sow {{Posture}} and {{Feeding Activity Monitoring}}
  in a {{Farrowing Pen}} Using {{Ground Vibration}}}. In
  \bibinfo{booktitle}{\emph{{{ECPLF}} 2022 - 10th {{European Conferenc}} on
  {{Precision Livestock Farming}}}} ({Vienna, Austria}, 2022-08-30).
\newblock


\bibitem[De~Jonge et~al\mbox{.}(1996)]%
        {de1996rearing}
\bibfield{author}{\bibinfo{person}{Francien~H De~Jonge}, \bibinfo{person}{EAM
  Bokkers}, \bibinfo{person}{WGP Schouten}, {and} \bibinfo{person}{FA
  Helmond}.} \bibinfo{year}{1996}\natexlab{}.
\newblock \showarticletitle{Rearing piglets in a poor environment:
  developmental aspects of social stress in pigs}.
\newblock \bibinfo{journal}{\emph{Physiology \& behavior}}
  \bibinfo{volume}{60}, \bibinfo{number}{2} (\bibinfo{year}{1996}),
  \bibinfo{pages}{389--396}.
\newblock


\bibitem[Dong et~al\mbox{.}(2023)]%
        {dong2023stranger}
\bibfield{author}{\bibinfo{person}{Yiwen Dong}, \bibinfo{person}{Jonathon
  Fagert}, \bibinfo{person}{Pei Zhang}, {and} \bibinfo{person}{Hae~Young Noh}.}
  \bibinfo{year}{2023}\natexlab{}.
\newblock \showarticletitle{Stranger Detection and Occupant Identification
  Using Structural Vibrations}. In \bibinfo{booktitle}{\emph{European Workshop
  on Structural Health Monitoring}}. Springer, \bibinfo{pages}{905--914}.
\newblock


\bibitem[Dong et~al\mbox{.}(2020)]%
        {dong2020md}
\bibfield{author}{\bibinfo{person}{Yiwen Dong}, \bibinfo{person}{Joanna~Jiaqi
  Zou}, \bibinfo{person}{Jingxiao Liu}, \bibinfo{person}{Jonathon Fagert},
  \bibinfo{person}{Mostafa Mirshekari}, \bibinfo{person}{Linda Lowes},
  \bibinfo{person}{Megan Iammarino}, \bibinfo{person}{Pei Zhang}, {and}
  \bibinfo{person}{Hae~Young Noh}.} \bibinfo{year}{2020}\natexlab{}.
\newblock \showarticletitle{MD-Vibe: physics-informed analysis of
  patient-induced structural vibration data for monitoring gait health in
  individuals with muscular dystrophy}. In \bibinfo{booktitle}{\emph{Adjunct
  proceedings of the 2020 ACM international joint conference on pervasive and
  ubiquitous computing and proceedings of the 2020 ACM international symposium
  on wearable computers}}. \bibinfo{pages}{525--531}.
\newblock


\bibitem[Jong et~al\mbox{.}(2000)]%
        {jong2000effects}
\bibfield{author}{\bibinfo{person}{IC~de Jong}, \bibinfo{person}{Andrea
  Sgoifo}, \bibinfo{person}{Elbert Lambooij}, \bibinfo{person}{S~Mechiel
  Korte}, \bibinfo{person}{Harry~J Blokhuis}, {and} \bibinfo{person}{Jaap~M
  Koolhaas}.} \bibinfo{year}{2000}\natexlab{}.
\newblock \showarticletitle{Effects of social stress on heart rate and heart
  rate variability in growing pigs}.
\newblock \bibinfo{journal}{\emph{Canadian Journal of Animal Science}}
  \bibinfo{volume}{80}, \bibinfo{number}{2} (\bibinfo{year}{2000}),
  \bibinfo{pages}{273--280}.
\newblock


\bibitem[Jorquera-Chavez et~al\mbox{.}(2020)]%
        {jorquera2020remotely}
\bibfield{author}{\bibinfo{person}{Maria Jorquera-Chavez},
  \bibinfo{person}{Sigfredo Fuentes}, \bibinfo{person}{Frank~R Dunshea},
  \bibinfo{person}{Robyn~D Warner}, \bibinfo{person}{Tomas Poblete},
  \bibinfo{person}{Rebecca~S Morrison}, {and} \bibinfo{person}{Ellen~C
  Jongman}.} \bibinfo{year}{2020}\natexlab{}.
\newblock \showarticletitle{Remotely sensed imagery for early detection of
  respiratory disease in pigs: a pilot study}.
\newblock \bibinfo{journal}{\emph{Animals}} \bibinfo{volume}{10},
  \bibinfo{number}{3} (\bibinfo{year}{2020}), \bibinfo{pages}{451}.
\newblock


\bibitem[Macon et~al\mbox{.}(2021)]%
        {Macon2021}
\bibfield{author}{\bibinfo{person}{Asya Macon}, \bibinfo{person}{Sudhendu
  Sharma}, \bibinfo{person}{Eric Markvicka}, \bibinfo{person}{Gary Rohrer},
  {and} \bibinfo{person}{Jeremy Miles}.} \bibinfo{year}{2021}\natexlab{}.
\newblock \showarticletitle{{Characterizing Lactating Sow Posture in Farrowing
  Crates Utilizing Automated Image Capture and Wearable Sensors}}.
  \bibinfo{pages}{634--642}.
\newblock


\bibitem[Marchant et~al\mbox{.}(2001)]%
        {marchant2001vocalisations}
\bibfield{author}{\bibinfo{person}{Jeremy~N Marchant}, \bibinfo{person}{Xanthe
  Whittaker}, {and} \bibinfo{person}{Donald~M Broom}.}
  \bibinfo{year}{2001}\natexlab{}.
\newblock \showarticletitle{Vocalisations of the adult female domestic pig
  during a standard human approach test and their relationships with
  behavioural and heart rate measures}.
\newblock \bibinfo{journal}{\emph{Applied Animal Behaviour Science}}
  \bibinfo{volume}{72}, \bibinfo{number}{1} (\bibinfo{year}{2001}),
  \bibinfo{pages}{23--39}.
\newblock


\bibitem[Marchant-Forde et~al\mbox{.}(2004)]%
        {MARCHANTFORDE2004449}
\bibfield{author}{\bibinfo{person}{R.M. Marchant-Forde}, \bibinfo{person}{D.J.
  Marlin}, {and} \bibinfo{person}{J.N. Marchant-Forde}.}
  \bibinfo{year}{2004}\natexlab{}.
\newblock \showarticletitle{Validation of a cardiac monitor for measuring heart
  rate variability in adult female pigs: accuracy, artefacts and editing}.
\newblock \bibinfo{journal}{\emph{Physiology and Behavior}}
  \bibinfo{volume}{80}, \bibinfo{number}{4} (\bibinfo{year}{2004}),
  \bibinfo{pages}{449--458}.
\newblock
\showISSN{0031-9384}
\urldef\tempurl%
\url{https://doi.org/10.1016/j.physbeh.2003.09.007}
\showDOI{\tempurl}


\bibitem[Mirshekari et~al\mbox{.}(2018)]%
        {mirshekari2018occupant}
\bibfield{author}{\bibinfo{person}{Mostafa Mirshekari}, \bibinfo{person}{Shijia
  Pan}, \bibinfo{person}{Jonathon Fagert}, \bibinfo{person}{Eve~M Schooler},
  \bibinfo{person}{Pei Zhang}, {and} \bibinfo{person}{Hae~Young Noh}.}
  \bibinfo{year}{2018}\natexlab{}.
\newblock \showarticletitle{Occupant localization using footstep-induced
  structural vibration}.
\newblock \bibinfo{journal}{\emph{Mechanical Systems and Signal Processing}}
  \bibinfo{volume}{112} (\bibinfo{year}{2018}), \bibinfo{pages}{77--97}.
\newblock


\bibitem[Randall(1990)]%
        {randall1990induction}
\bibfield{author}{\bibinfo{person}{GC Randall}.}
  \bibinfo{year}{1990}\natexlab{}.
\newblock \showarticletitle{Induction of parturition in pigs: short term
  effects of prostaglandin F2 alpha on chronically catheterised fetuses at
  term.}
\newblock \bibinfo{journal}{\emph{The Veterinary Record}}
  \bibinfo{volume}{126}, \bibinfo{number}{3} (\bibinfo{year}{1990}),
  \bibinfo{pages}{61--63}.
\newblock


\bibitem[Sipos et~al\mbox{.}(2013)]%
        {sipos2013physiological}
\bibfield{author}{\bibinfo{person}{W Sipos}, \bibinfo{person}{S Wiener},
  \bibinfo{person}{F Entenfellner}, \bibinfo{person}{S Sipos}, {et~al\mbox{.}}}
  \bibinfo{year}{2013}\natexlab{}.
\newblock \showarticletitle{Physiological changes of rectal temperature, pulse
  rate and respiratory rate of pigs at different ages including the critical
  peripartal period}.
\newblock \bibinfo{journal}{\emph{Vet. Med. Austria}} \bibinfo{volume}{100},
  \bibinfo{number}{3} (\bibinfo{year}{2013}), \bibinfo{pages}{96}.
\newblock


\bibitem[Smith(1956)]%
        {smith1956respiration}
\bibfield{author}{\bibinfo{person}{Thomas~C Smith}.}
  \bibinfo{year}{1956}\natexlab{}.
\newblock \showarticletitle{The respiration and composition of the mammary
  gland of the guinea pig during pregnancy and lactation}.
\newblock \bibinfo{journal}{\emph{Archives of Biochemistry and Biophysics}}
  \bibinfo{volume}{60}, \bibinfo{number}{2} (\bibinfo{year}{1956}),
  \bibinfo{pages}{485--495}.
\newblock


\bibitem[{von Borell} et~al\mbox{.}(2007)]%
        {VONBORELL2007293}
\bibfield{author}{\bibinfo{person}{Eberhard {von Borell}}, \bibinfo{person}{Jan
  Langbein}, \bibinfo{person}{Gérard Després}, \bibinfo{person}{Sven Hansen},
  \bibinfo{person}{Christine Leterrier}, \bibinfo{person}{Jeremy
  Marchant-Forde}, \bibinfo{person}{Ruth Marchant-Forde},
  \bibinfo{person}{Michela Minero}, \bibinfo{person}{Elmar Mohr},
  \bibinfo{person}{Armelle Prunier}, \bibinfo{person}{Dorothée Valance}, {and}
  \bibinfo{person}{Isabelle Veissier}.} \bibinfo{year}{2007}\natexlab{}.
\newblock \showarticletitle{Heart rate variability as a measure of autonomic
  regulation of cardiac activity for assessing stress and welfare in farm
  animals — A review}.
\newblock \bibinfo{journal}{\emph{Physiology and Behavior}}
  \bibinfo{volume}{92}, \bibinfo{number}{3} (\bibinfo{year}{2007}),
  \bibinfo{pages}{293--316}.
\newblock
\showISSN{0031-9384}
\urldef\tempurl%
\url{https://doi.org/10.1016/j.physbeh.2007.01.007}
\showDOI{\tempurl}
\newblock
\shownote{Stress and Welfare in Farm Animals}.


\bibitem[Wang et~al\mbox{.}(2021)]%
        {wang2021contactless}
\bibfield{author}{\bibinfo{person}{Meiqing Wang}, \bibinfo{person}{Ali
  Youssef}, \bibinfo{person}{Mona Larsen}, \bibinfo{person}{Jean-Loup Rault},
  \bibinfo{person}{Daniel Berckmans}, \bibinfo{person}{Jeremy~N
  Marchant-Forde}, \bibinfo{person}{Joerg Hartung}, \bibinfo{person}{Andr{\'e}
  Bleich}, \bibinfo{person}{Mingzhou Lu}, {and} \bibinfo{person}{Tomas
  Norton}.} \bibinfo{year}{2021}\natexlab{}.
\newblock \showarticletitle{Contactless video-based heart rate monitoring of a
  resting and an anesthetized pig}.
\newblock \bibinfo{journal}{\emph{Animals}} \bibinfo{volume}{11},
  \bibinfo{number}{2} (\bibinfo{year}{2021}), \bibinfo{pages}{442}.
\newblock


\bibitem[Zaleski and Hacker(1993)]%
        {zaleski1993variables}
\bibfield{author}{\bibinfo{person}{Halina~M Zaleski} {and}
  \bibinfo{person}{Roger~R Hacker}.} \bibinfo{year}{1993}\natexlab{}.
\newblock \showarticletitle{Variables related to the progress of parturition
  and probability of stillbirth in swine}.
\newblock \bibinfo{journal}{\emph{The Canadian Veterinary Journal}}
  \bibinfo{volume}{34}, \bibinfo{number}{2} (\bibinfo{year}{1993}),
  \bibinfo{pages}{109}.
\newblock


\bibitem[{\v{Z}}Emaityt{\.e} et~al\mbox{.}(1984)]%
        {vzemaityte1984heart}
\bibfield{author}{\bibinfo{person}{Danguol{\.e} {\v{Z}}Emaityt{\.e}},
  \bibinfo{person}{Giedrius Varoneckas}, {and} \bibinfo{person}{Eugene
  Sokolov}.} \bibinfo{year}{1984}\natexlab{}.
\newblock \showarticletitle{Heart rhythm control during sleep}.
\newblock \bibinfo{journal}{\emph{Psychophysiology}} \bibinfo{volume}{21},
  \bibinfo{number}{3} (\bibinfo{year}{1984}), \bibinfo{pages}{279--289}.
\newblock


\end{thebibliography}

\end{document}